\documentclass[fleqn,12pt,twoside]{article}
\usepackage[headings]{espcrc1}

\readRCS
$Id: espcrc1.tex,v 1.2 2004/02/24 11:22:11 spepping Exp $
\usepackage{graphicx}
\usepackage[figuresright]{rotating}
\begin{document}
\newcommand{\ttbs}{\char'134}
\newcommand{\AmS}{{\protect\the\textfont2
  A\kern-.1667em\lower.5ex\hbox{M}\kern-.125emS}}

\hyphenation{author another created financial paper re-commend-ed Post-Script}

\title{Isospin Dynamics in Heavy Ion Collisions: EoS-sensitive Observables}
\author{M.Di Toro\address[lns]{Laboratori Nazionali del Sud INFN, I-95123 
Catania, Italy,\\
and Physics-Astronomy Dept., University of Catania}%
\thanks{ditoro@lns.infn.it},
V.Baran\address[bucharest]{Dept.of Theoretical Physics, Bucharest Univ., 
Magurele, Bucharest, Romania},
M.Colonna\addressmark[lns],
G.Ferini\addressmark[lns],
T.Gaitanos\address[muenchen]{Dept. f\"ur Physik, Universit\"at M\"unchen, 
D-85748 Garching, Germany},
V.Greco\addressmark[lns],
J.Rizzo\addressmark[lns],
H.H.Wolter\addressmark[muenchen].}

\runtitle{Isospin Dynamics}
\runauthor{M.Di Toro}

\maketitle

\begin{abstract}
Heavy Ion Collisions ($HIC$) represent a unique tool to probe the in-medium
nuclear interaction in regions away from saturation and at high nucleon
momenta. In this report we present a selection of reaction observables 
particularly sensitive to the isovector part of the interaction, i.e. to the
symmetry term of the nuclear Equation of State ($EoS$)
At low energies the behavior of the symmetry energy around saturation 
influences dissipation and fragment production mechanisms.
Predictions 
are shown for deep-inelastic and fragmentation collisions induced by neutron 
rich projectiles. Differential flow measurements will also shed lights 
on the controversial neutron/proton effective mass splitting in asymmetric 
matter. The high density symmetry term can be derived from
 isospin effects on heavy ion reactions
at relativistic energies (few $AGeV$ range), that can even allow
a ``direct'' study of the covariant structure of the isovector interaction
in the hadron medium. 
Rather sensitive observables are proposed from collective flows
and from pion/kaon production. 
The possibility of the transition to a mixed hadron-quark phase, 
at high baryon and isospin density, is finally suggested. Some signatures
could come from an expected ``neutron trapping'' effect.

\end{abstract}

\vskip -1.0cm
\section{Introduction}

The symmetry energy $E_{sym}$ appears in the energy density
$\epsilon(\rho,\rho_3) \equiv \epsilon(\rho)+\rho E_{sym} (\rho_3/\rho)^2
 + O(\rho_3/\rho)^4 +..$, expressed in terms of total ($\rho=\rho_p+\rho_n$)
 and isospin ($\rho_3=\rho_p-\rho_n$) densities. The symmetry term gets a
kinetic contribution directly from basic Pauli correlations and a potential
part from the highly controversial isospin dependence of the effective 
interactions \cite{baranPR}. Both at sub-saturation and supra-saturation
densities, predictions based of the existing many-body techniques diverge 
rather widely, see \cite{fuchswci}. 
We take advantage of new opportunities in 
theory (development of rather reliable microscopic transport codes for $HIC$)
 and in experiments (availability of very asymmetric radioactive beams, 
improved possibility of measuring event-by-event correlations) to present
results that are severely constraining the existing effective interaction 
models. We will discuss dissipative collisions in a wide range of energies,
 from just above the Coulomb barrier up to a few $AGeV$. 
The transport codes are based on 
mean field theories, with correlations included via hard nucleon-nucleon
elastic and inelastic collisions and via stochastic forces, selfconsistently
evaluated from the mean phase-space trajectory, see 
\cite{baranPR,guarneraPLB373,colonnaNPA642,chomazPR}. Stochasticity is 
essential in 
order to get distributions as well as to allow the growth of dynamical 
instabilities.  

\section{Isospin effects on Deep-Inelastic Collisions}
\begin{figure}
\vskip 1.0cm
\begin{center}
 \includegraphics[scale=0.40]{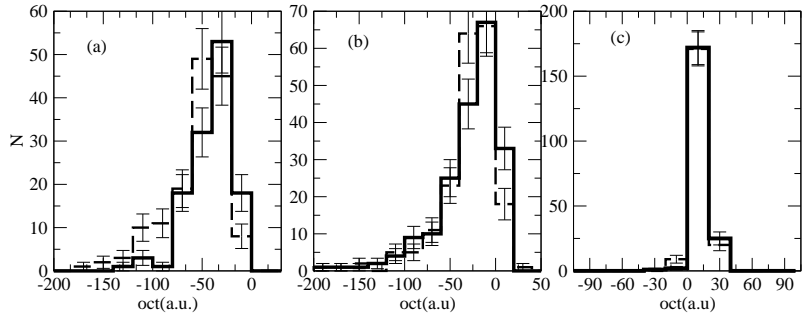}
\vskip -0.5cm
\caption{Distribution of the octupole moment of primary fragments for 
the $^{132}Sn+^{64}Ni$
reaction at $10~AMeV$ (impact parameters (a):$b=6fm$, (b):$7fm$, (c):$8fm$).
Solid lines: asysoft. Dashed lines: asystiff} 
\vskip -1.3cm
\label{octupole}
\end{center}
\end{figure}

Dissipative semi-peripheral collisions at low energies, including binary and 
three-body breakings, offer a good opportunity to study phenomena occurring 
in nuclear matter under extreme conditions with respect to shape, excitation 
energy, spin and N/Z ratio (isospin). 
In some cases, due to a combined Coulomb and angular momentum 
(deformation) effect, some instabilities can show up \cite{colonnaNPA589}.
 This can lead to 3-body breakings, where a light cluster is emitted from the
 neck region. Three body processes in collisions with exotic beams will allow 
to investigate how the development of surface (neck-like) instabilities, that 
would help ternary breakings, is sensitive to the structure of the symmetry 
term around (below) saturation.
In order to suggest proposals for the new $RIB$ facility $Spiral~2$, 
\cite{lewrio} we have studied the reaction $^{132}Sn+^{64}Ni$ at $10AMeV$
in semicentral events, impact parameters $b=6, 7, 8 fm$, where one observes 
mostly binary exit channels, but still in presence of large dissipation. Two 
different behaviors of the symmetry energy below saturation have been tested: 
one ($asysoft$) where it is a smooth decreasing function towards low densities,
 and another one ($asystiff$) where we have a rapid decrease, \cite{baranPR}.
The Wilczynski plots, kinetic energy loss vs. deflection angle, show slightly 
more dissipative events in the $asystiff$ case, consistent with the point that 
in the interaction at lower densities in very neutron-rich matter (the neck 
region) we have a less repulsive symmetry term. In fact the 
neck dynamics is 
rather different in the two cases, as it can be well  evidenced looking at the
 deformation of the $PLF/TLF$ residues. The distribution of the octupole 
moment over the considered ensemble of events is shown in Fig.\ref{octupole} 
for the three considered impact parameters.
Except for the most peripheral events, larger deformations, strongly 
suggesting a final 3-body outcome, are seen in the $asystiff$ case. Now, 
due to the lower value of the symmetry enrgy, the neutron-rich neck 
connecting the two systems survives a longer time leading to very deformed 
primary fragments, from which eventually small clusters will be dynamically 
emitted.
Finally we expect to see effects of the different interaction times on the 
charge equilibration mechanism, probed starting from entrance channels with 
large $N/Z$ asymmetries, like $^{132}Sn(N/Z=1.64)+^{58}Ni(N/Z=1.07)$. Moreover 
the equilibration mechanism is also directly driven by the strenght of the 
symmetry term.

\section{Isospin Dynamics in Neck Fragmentation at Fermi Energies}

It is now quite well established that the largest part of the reaction
cross section for dissipative collisions at Fermi energies goes
through the {\it Neck Fragmentation} channel, with $IMF$s directly
produced in the interacting zone in semiperipheral collisions on very short
time scales \cite{wcineck}. We can predict interesting isospin transport 
effects for this new
fragmentation mechanism since clusters are formed still in a dilute
asymmetric matter but always in contact with the regions of the
projectile-like and target-like remnants almost at normal densities.
Since the difference between local neutron-proton chemical potentials is given 
by $\mu_n-\mu_p=4E_{sym}(\rho_3/\rho)$, we expect a larger neutron flow to
 the neck clusters for a stiffer symmetry energy around saturation, 
\cite{baranPR,baranPRC72}. The isospin dynamics can be directly extracted 
from correlations between $N/Z$, $alignement$ and emission times of the $IMF$s.
The alignment between $PLF-IMF$ and $PLF-TLF$ directions
represents a very convincing evidence of the dynamical origin of the 
mid-rapidity fragments produced on short time scales \cite{baranNPA730}. 
The form of the
$\Phi_{plane}$ distributions (centroid and width) can give a direct
information on the fragmentation mechanism \cite{dynfiss05}. Recent 
calculations confirm that the light fragments are emitted first, a general 
feature expected for that rupture mechanism \cite{liontiPLB625}. 
The same conclusion can be derived from {\it direct} emission time 
measurements based on deviations from Viola systematics  observed
in event-by-event velocity correlations between $IMF$s and the $PLF/TLF$ 
residues
 \cite{baranNPA730,dynfiss05,velcorr04}. 
 We can figure out
   a continuous transition from fast produced fragments via neck instabilities
   to clusters formed in a dynamical fission of the projectile(target) 
   residues up to the evaporated ones (statistical fission). Along this 
   line it would be even possible to disentangle the effects of volume
   and shape instabilities. 
A neutron enrichment of the overlap ("neck") region is
   expected, due to the neutron migration from higher (spectator) to 
   lower (neck) density regions, directly related to
   the slope of the symmetry energy \cite{liontiPLB625}. 
A very nice new analysis has been presented on the $Sn+Ni$ data at $35~AMeV$
by the Chimera Collab., Fig.2 of ref.\cite{defilposter}.
A strong correlation between neutron enrichemnt and alignement (when the 
short emission time selection is enforced) is seen, that can be reproduced 
only with 
a stiff behavior of the symmetry energy. {\it This is the 
first clear evidence in favor of a relatively large slope (symmetry pressure) 
around saturation}.

\section{Effective Mass Splitting and Collective Flows}

The problem of Momentum Dependence in the Isovector
channel ($Iso-MD$) is still very controversial and it would be extremely
important to get more definite experimental information,
see the recent refs. 
\cite{BaoNPA735,RizzoNPA732,ditoroAIP05,rizzoPRC72,ZuoPRC72,DalenarPRL95}. 
Intermediate energies are
important in order to have high momentum particles and to test regions
of high baryon (isoscalar) and isospin (isovector) density during the
reactions dynamics.
Collective flows \cite{OlliPRD46} are very good candidates since they are 
expected to be 
very sensitive to the momentum
dependence of the mean field, see \cite{DanielNPA673,baranPR}.
The transverse flow, 
$V_1(y,p_t)=\langle \frac{p_x}{p_t} \rangle$,
provides information on the anisotropy of 
nucleon emission on the reaction plane.
Very important for the reaction dynamics is the elliptic
flow,
$V_2(y,p_t)=\langle \frac{p_x^2-p_y^2}{p_t^2} \rangle$.
 The sign of $V_2$ indicates the azimuthal anisotropy of emission:
on the reaction
plane ($V_2>0$) or out-of-plane ($squeeze-out,~V_2<0$)
\cite{OlliPRD46,DanielNPA673}.
We have then tested the $Iso-MD$ of the fields
just evaluating the $Difference$ of neutron/proton transverse and elliptic 
flows 
$
V^{(n-p)}_{1,2} (y,p_t) \equiv V^n_{1,2}(y,p_t) - V^p_{1,2}(y,p_t)
$ 
at various rapidities and transverse momenta in semicentral
($b/b_{max}=0.5$) $^{197}Au+^{197}Au$ collisons at $250AMeV$, where
some proton data are existing from the $FOPI$ collaboration at $GSI$
 \cite{fopi_v1,fopi_v2}.
The transport code has been implemented with 
a $BGBD-like$ \cite{GalePRC41,BombaciNPA583}   mean field 
with a different $(n,p)$ momentum dependence, see 
\cite{RizzoNPA732,ditoroAIP05,rizzoPRC72}, that allow to follow the dynamical
effect of opposite n/p effective mass splitting while keeping the
same density dependence of the symmetry energy.
\begin{figure}[htb]
\centering
\begin{picture}(0,0)
\put(150.0,4.8){\mbox{\includegraphics[width=4.2cm]{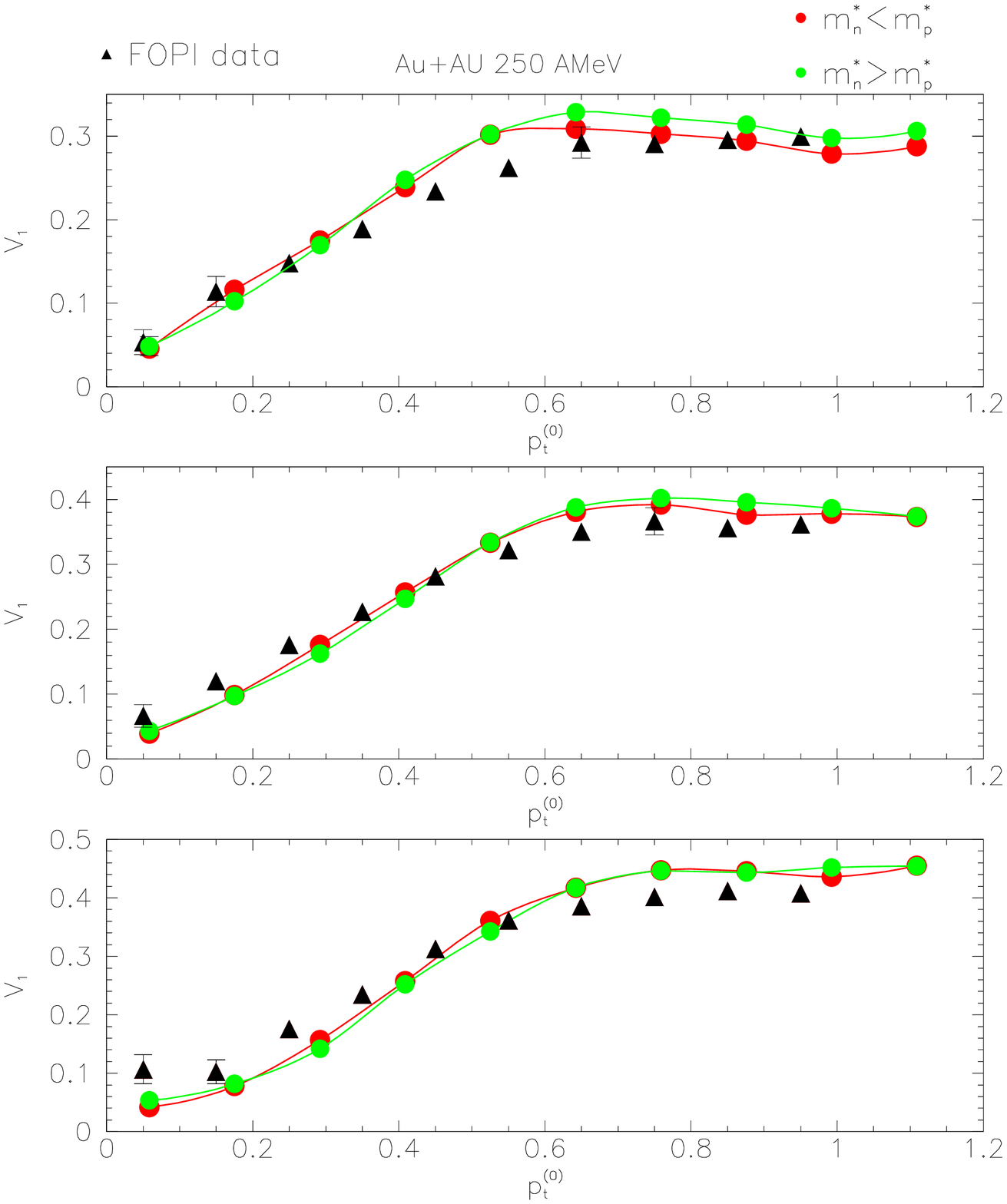}}}
\end{picture}
\includegraphics[width=9cm]{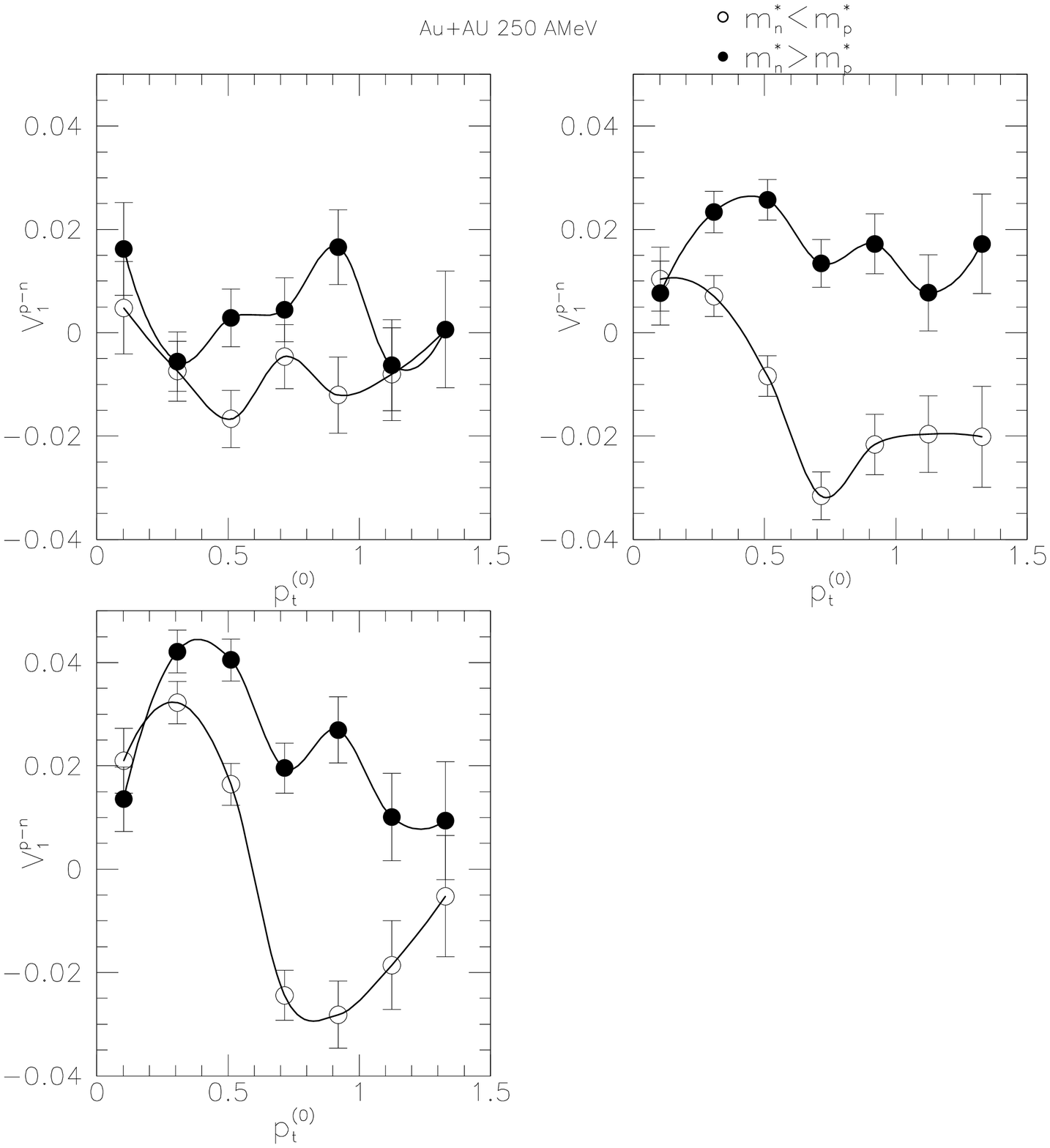}
\vskip -1.0cm
\caption{Difference between proton and neutron $V_1$ flows in a 
semi-central
reaction Au+Au at 250\, AMeV for three rapidity ranges. Upper Left Panel:
$|y^{(0)}| \leq 0.3$; Upper Right: $0.3 \leq|y^{(0)}| \leq 0.7$;
Lower Left: $0.6 \leq|y^{(0)}| \leq 0.9$.
Lower Right Panel: Comparison of the $V_1$ proton flow 
with FOPI data \cite{fopi_v1} 
for three rapidity ranges. Top: $0.5 \leq|y^{(0)}| \leq 0.7$;
center: $0.7 \leq|y^{(0)}| \leq 0.9$; 
bottom: $0.9 \leq|y^{(0)}| \leq 1.1$.
} 
\vskip -1.0cm
\label{v1dif}
\end{figure}

For the difference of nucleon transverse flows, see Fig. \ref{v1dif}, 
 the mass splitting
effect is evident at all rapidities, and nicely increasing at larger
rapidities and transverse momenta, with more neutron flow when
$m_n^*<m_p^*$.
Just to show that our simulations give realistic results we compare in 
lower right panel of Fig. \ref{v1dif} with the proton data of 
the $FOPI$ collaboration
for similar selections of impact parameters rapidities and transverse momenta.
The same analysis has been performed for the difference of elliptic flows,
\cite{ditoroAIP05}. Again the mass splitting effects are more evident
for higher rapidity and tranverse momentum selections. In particular
the differential elliptic flow becomes negative when
$m_n^*<m_p^*$, revealing a faster neutron emission and so more neutron
squeeze out (more spectator shadowing).
The measurement of n/p flow differences appears essential.
Due to the difficulties in
measuring neutrons, our suggestion is to measure the difference between
light isobar flows, like $^3H$ vs. $^3He$ and so on.
We expect to clearly see the effective mass splitting effects, 
 maybe even enhanced due to larger overall flows shown by clusters, see 
 \cite{baranPR,ScalonePLB461}.
\vskip -1.0cm

\section{Relativistic Collisions}
 Finally we focus our attention on relativistic heavy ion collisions, that
provide a unique terrestrial opportunity to probe the in-medium nuclear
interaction at high densities. 
An effective Lagrangian approach to the hadron interacting system is
extended to the isospin degree of freedom: within the same frame equilibrium
properties ($EoS$, \cite{qhd}) and transport dynamics 
\cite{KoPRL59,GiessenRPP56} can be consistently derived.
Within a covariant picture of the nuclear mean field, 
 for the description of the symmetry energy at saturation
($a_{4}$ parameter of the Weizs\"{a}ecker mass formula)
(a) only the Lorentz vector $\rho$ mesonic field, 
and (b) both, the vector $\rho$ (repulsive) and  scalar 
$\delta$ (attractive) effective 
fields \cite{liu,gait04} can be included. 
In the latter case the competition between scalar and vector fields leads
to a stiffer symmetry term at high density \cite{liu,baranPR}. 
The presence of the hadronic medium leads to effective masses and 
momenta $M^{*}=M+\Sigma_{s}$,   
 $k^{*\mu}=k^{\mu}-\Sigma^{\mu}$, with
$\Sigma_{s},~\Sigma^{\mu}$
 scalar and vector self-energies. 
For asymmetric matter the self-energies are different for protons and 
neutrons, depending on the isovector meson contributions. 
We will call the 
corresponding models as $NL\rho$ and $NL\rho\delta$, respectively, and
just $NL$ the case without isovector interactions. 

For the description of heavy ion collisions we solve
the covariant transport equation of the Boltzmann type 
 \cite{KoPRL59,GiessenRPP56}  within the 
Relativistic Landau
Vlasov ($RLV$) method, using phase-space Gaussian test particles 
\cite{FuchsNPA589},
and applying
a Monte-Carlo procedure for the hard hadron collisions.
The collision term includes elastic and inelastic processes involving
the production/absorption of the $\Delta(1232 MeV)$ and $N^{*}(1440
MeV)$ resonances as well as their decays into pion channels,
 \cite{FeriniNPA762,feriniarxiv}.
A larger repulsive vector contribution to the neutron energies is given by the
$\rho$-coupling. This is rapidly increasing with density when the $\delta$ 
field is included \cite{liu,baranPR}. As a consequence we expect a good 
sensitivity to the covariant structure of the isovector fields in nucleon 
emission and particle production data. Moreover the presence of a 
{\it Lorentz magnetic} term in the relativistic transport equation
\cite{KoPRL59,GiessenRPP56,baranPR} will enhance the dynamical effects
of vector fields \cite{GrecoPLB562}.

Differential flows will be directly affected.
\begin{figure}
\begin{center}
 \includegraphics[angle=-90,scale=0.35]{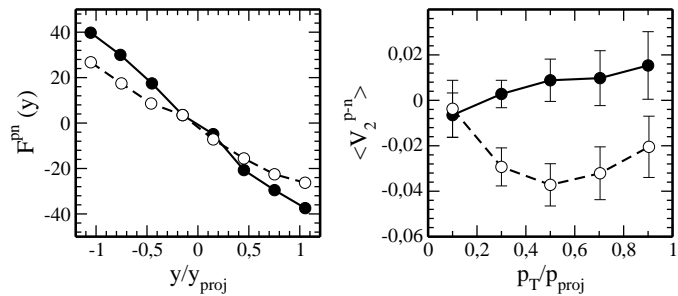}
\vskip -1.0cm
\caption{\small{Differential neutron-proton flows for the $^{132}Sn+^{124}Sn$
reaction at $1.5~AGeV$ ($b=6fm$) from the two different models for the
isovector mean fields.
Left: Transverse Flows. Right: Elliptic Flows.
Full circles and solid line: $NL\rho\delta$.
Open circles and dashed line: $NL\rho$.}
} 
\vskip -1.3cm
\label{flows}
\end{center}
\end{figure}
In Fig.\ref{flows}
transverse and elliptic differential flows are shown
for 
the $^{132}Sn+^{124}Sn$
reaction at $1.5~AGeV$ ($b=6fm$), 
 \cite{GrecoPLB562}. 
The effect of the different structure of the 
isovector channel is clear. Particularly evident is the splitting in 
the high $p_t$
region of the elliptic flow.
 In the $(\rho+\delta)$ dynamics the high-$p_t$ neutrons show a much larger 
$squeeze-out$.
This is fully consistent with an early emission (more spectator shadowing)
due to the larger $\rho$-field in the compression stage.

\vskip -1.0cm
\section{Isospin effects on sub-threshold kaon production at intermediate 
energies} 
Kaon production has been proven to be a reliable observable for the
high density $EoS$ in the isoscalar sector 
\cite{FuchsPPNP56,HartPRL96}
Here we show that the $K^{0,+}$
production (in particular the $K^0/K^+$ yield ratio) can be also used to 
probe the isovector part of the $EoS$.

Using our $RMF$ transport approach  we analyze 
pion and kaon production in central $^{197}Au+^{197}Au$ collisions in 
the $0.8-1.8~AGeV$
 beam 
energy range, comparing models giving the same ``soft'' $EoS$ for symmetric 
matter and with different effective field choices for 
$E_{sym}$ \cite{feriniarxiv}. Here we also use a Lagrangian
with density
dependent couplings ($DDF$, \cite{gait04}), recently suggested 
for better nucleonic properties of neutron stars \cite{Klahn06}.
In the $DDF$ model
the  $\rho$-coupling is exponentially decreasing with density, resulting in a 
rather "soft" 
symmetry term at high density. 
The hadron mean field propagation, which goes beyond the 
``collision cascade'' picture, is
essential for particle production yields: in particular the
isospin dependence of the self-energies directly affects the
energy balance of the inelastic channels.

\begin{figure}[t] 
\begin{center}
\includegraphics[scale=0.28]{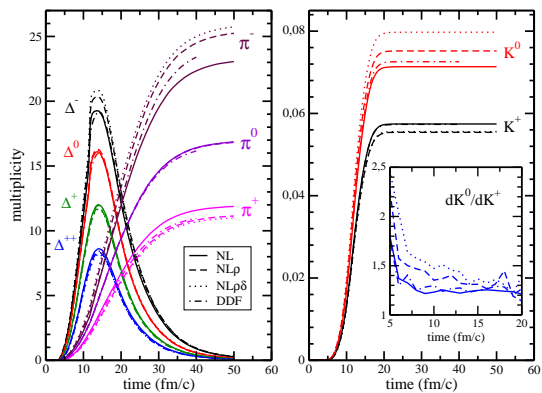} 
\vskip -1.0cm
\caption{\small{Time evolution of the $\Delta^{\pm,0,++}$ resonances
and pions $\pi^{\pm,0}$ 
(left),  and  kaons ($K^{+,0}$
 (right) for a central ($b=0$ fm impact parameter)  
Au+Au collision at 1 AGeV incident energy. Transport calculation using the  
$NL, NL\rho, NL\rho\delta$ and $DDF$ models for the iso-vector part of the  
nuclear $EoS$ are shown.}  
}
\vskip -1.3cm
\label{kaon1} 
\end{center}
\end{figure}

Fig. \ref{kaon1} reports  the temporal evolution of $\Delta^{\pm,0,++}$  
resonances, pions ($\pi^{\pm,0}$) and kaons ($K^{+,0}$)  
for central Au+Au collisions at $1AGeV$.
It is clear that, while the pion yield freezes out at times of the order of 
$50 fm/c$, i.e. at the final stage of the reaction (and at low densities),
kaon production occur within the very early (compression) stage,
 and the yield saturates at around $20 fm/c$. 
From Fig. \ref{kaon1} we see that the pion results are  
weakly dependent on the  
isospin part of the nuclear mean field.
However, a slight increase (decrease) in the $\pi^{-}$ ($\pi^{+}$) 
multiplicity is observed when going from the $NL$ (or $DDF$) to the 
$NL\rho$ and then to
the $NL\rho\delta$ model, i.e. increasing the vector contribution $f_\rho$
in the isovector channel. This trend is 
more pronounced for kaons, see the
right panel, due to the high density selection of the source and the
proximity to the production threshold. 

When isovector fields are included the symmetry potential energy in 
neutron-rich matter is repulsive for neutrons and attractive for protons.
In a $HIC$ this leads to a fast, pre-equilibrium, emission of neutrons.
 Such a $mean~field$ mechanism, often referred to as isospin fractionation
\cite{baranPR}, is responsible for a reduction of the neutron
to proton ratio during the high density phase, with direct consequences
on particle production in inelastic $NN$ collisions.
$Threshold$ effects represent a more subtle point. The energy 
conservation in
a hadron collision is expressed in terms of the canonical
momenta, i.e. for a reaction $1+2 \rightarrow 3+4$ as
$
s_{in} = (k_1^\mu + k_2^\mu)^2 = (k_3^\mu + k_4^\mu)^2 = s_{out}.
$
Since hadrons are propagating with effective (kinetic) momenta and masses,
 an equivalent relation should be formulated starting from the effective
in-medium quantities $k^{*\mu}=k^\mu-\Sigma^\mu$ and $m^*=m+\Sigma_s$, where
$\Sigma_s$ and $\Sigma^\mu$ are the scalar and vector self-energies.
The self-energy contributions will influence the particle production at the
level of thresholds as well as of the phase space available in the final 
channel.
In neutron-rich colliding systems {\it Mean field} 
and {\it threshold} effects
are acting in opposite directions. At low energies, around the production
threshold,, the energy conservation (i.e. the self energy contributions)
 is dominant, as we see from Fig. \ref{kaon1}, in particular for kaons.

We have to note that in a previous study of kaon production in excited nuclear
matter the dependence of the $K^{0}/K^{+}$ yield ratio on the effective
isovector interaction appears much larger (see Fig.8 of 
ref.\cite{FeriniNPA762}).
The point is that in the non-equilibrium case of a heavy ion collision
the asymmetry of the source where kaons are produced is in fact reduced
by the $n \rightarrow p$ ``transformation'', due to the favored 
$nn \rightarrow p\Delta^-$ processes. This effect is almost absent at 
equilibrium due to the inverse transitions.
Moreover in infinite nuclear matter even the fast
neutron emission is not present. 
This result clearly shows that chemical equilibrium models can lead to
uncorrect results when used for transient states of an $open$ system.

\section{Testing Deconfinement at High Isospin Density}
The hadronic matter is expected to undergo a phase transition 
into a deconfined phase of quarks and gluons at large densities 
and/or high temperatures. 
On very general grounds,
the transition critical densities are expected to depend
on the isospin of the system, but no experimental tests of this 
dependence have been performed so far.
In order to check the possibility of observing some precursor signals
of some new physics even in collisions of stable nuclei at
intermediate energies we have performed some event simulations for the
collision of very heavy, neutron-rich, elements. We have chosen the
reaction $^{238}U+^{238}U$ (average proton fraction $Z/A=0.39$) at
$1~AGeV$ and semicentral impact parameter $b=7~fm$ just to increase
the neutron excess in the interacting region. 
After about $10~fm/c$ in the overlap region a nice local
equilibration is achieved.  
A rather exotic nuclear matter is formed in a transient
time of the order of $10~fm/c$, with baryon density around $3-4\rho_0$,
temperature $50-60~MeV$, energy density $\approx 500~MeV~fm^{-3}$ and proton
fraction between $0.35$ and $0.40$, likely inside the estimated mixed 
phase region \cite{deconf06}.

We can study the isospin dependence of the transition densities
 \cite{MuellerNPA618} in a systematic way.
Concerning the hadronic phase, we use the relativistic
non-linear model of Glendenning-Moszkowski (in particular the ``soft''
$GM3$ choice) 
\cite{GlendenningPRL18}, where the isovector part is treated 
just with $\rho$ meson coupling, and
the iso-stiffer $NL\rho\delta$ interaction \cite{deconf06}. 
For the quark phase we consider the $MIT$ bag model 
with various bag pressure constants.  In particular 
we are interested in those parameter sets
which would allow the existence of quark stars
\cite{DragoPLB511}, i.e. parameter sets for
which the so-called Witten-Bodmer hypothesis is satisfied
\cite{WittenPRD30,BodmerPRD4}. 
One of the
aim of our work it to show that if quark stars are indeed possible,
it is then very likely to find signals of the formation of a mixed
quark-hadron phase in intermediate-energy heavy-ion experiments
\cite{deconf06}.
The structure of the mixed phase is obtained by
imposing the Gibbs conditions \cite{GlendenningPRD46} for
chemical potentials and pressure and by requiring
the conservation of the total baryon and isospin densities
\begin{eqnarray}\label{gibbs}
&&\mu_B^{(H)} = \mu_B^{(Q)}\, ,~~  
\mu_3^{(H)} = \mu_3^{(Q)} \, ,  
~~~P^{(H)}(T,\mu_{B,3}^{(H)}) = P^{(Q)} (T,\mu_{B,3}^{(Q)})\, ,\nonumber \\
&&\rho_B=(1-\chi)\rho_B^H+\chi\rho_B^Q \, ,
~~\rho_3=(1-\chi)\rho_3^H+\chi\rho_3^Q\, , 
\end{eqnarray}
where $\chi$ is the fraction of quark matter in the mixed phase.
In this way we get the $binodal$ surface which gives the phase coexistence 
region
in the $(T,\rho_B,\rho_3)$ space
\cite{GlendenningPRD46,MuellerNPA618}. For a fixed value of the
conserved charge $\rho_3$ 
 we will study the boundaries of the mixed phase
region in the $(T,\rho_B)$ plane. 
In the hadronic phase the charge chemical potential is given by
$
\mu_3 = 2 E_{sym}(\rho_B) \frac{\rho_3}{\rho_B}\, .
$ 
Thus, we expect critical densities
rather sensitive to the isovector channel in the hadronic $EoS$.

\begin{figure}
\begin{center}
\includegraphics[angle=+90,scale=0.37]{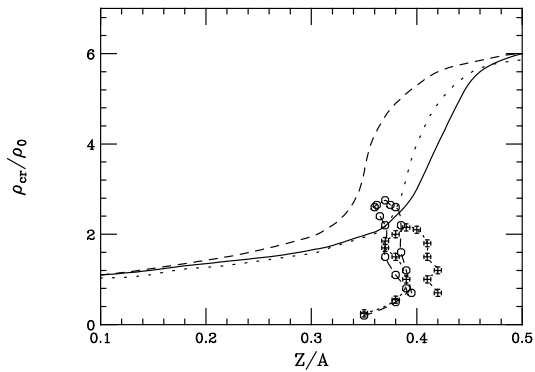}
\vskip -1.0cm
\caption{
\small{Variation of the transition density with proton fraction for various
hadronic $EoS$ parameterizations. Dotted line: $GM3$ parametrization;
 dashed line: $NL\rho$ parametrization; solid line: $NL\rho\delta$ 
parametrization. For the quark $EoS$, the $MIT$ bag model with
$B^{1/4}$=150 $MeV$.
The points represent the path followed
in the interaction zone during a semi-central $^{132}$Sn+$^{132}$Sn
collision at $1~AGeV$ (circles) and at $300~AMeV$ (crosses). 
}}
\vskip -1.3cm
\label{rhodelta}
\end{center}
\end{figure}

In Fig.~\ref{rhodelta}  we show the crossing
density $\rho_{cr}$ separating nuclear matter from the quark-nucleon
mixed phase, as a function of the proton fraction $Z/A$.  
We can see the effect of the
$\delta$-coupling towards an $earlier$ crossing due to the larger
symmetry repulsion at high baryon densities.
In the same figure we report the paths in the $(\rho,Z/A)$
plane followed in the c.m. region during the collision of the n-rich
 $^{132}$Sn+$^{132}$Sn system, at different energies. At
$300~AMeV$ we are just reaching the border of the mixed phase, and we are
well inside it at $1~AGeV$. 
Statistical
fluctuations could help in  reducing the density at which drops of quark
matter form. The reason is that a small bubble can
be energetically favored if it contains quarks whose Z/A ratio is
{\it smaller} than the average value of the surrounding region
 \cite {deconf06}. 
This corresponds to a {\it neutron trapping}
effect, supported also by a symmetry energy difference in the
two phases.
In fact while in the hadron phase we have a large neutron
potential repulsion (in particular in the $NL\rho\delta$ case), in the
quark phase we only have the much smaller kinetic contribution.
 If in a
pure hadronic phase neutrons are quickly emitted or ``transformed'' in
protons by inelastic collisions, when the mixed phase
starts forming, neutrons are kept in the interacting system up to the
subsequent hadronization in the expansion stage \cite{deconf06}.
Observables related to such neutron ``trapping'' could be an
inversion in the trend of the formation of neutron rich fragments
and/or of the $\pi^-/\pi^+$, $K^0/K^+$ yield ratios for reaction
products coming from high density regions, i.e. with large transverse
momenta.  

\section{Perspectives}
We have shown that {\it violent} collisions of n-rich heavy ions 
from low to relativistic energies
can bring new information on the isovector part of the in-medium interaction, 
qualitatively different from equilibrium
$EoS$ properties. We have presented quantitative results 
in a wide range of beam energies.
At low energies we have shown isospin effects on the dissipation in deep 
inelastic collisions, at Fermi energies the Iso-EoS sensitivity of the isospin 
transport in fragment reactions and finally at intermediate the dependence of
differential flows on the $Iso-MD$ and effective mass splitting. 
In relativistic collisions we have shown the possibility of a direct 
$measure$ of the Lorentz structure of the isovector fields at high baryon 
density, from
differential
collective flows and yields of charged pion and kaon ratios.
Important non-equilibrium effects for particle production are stressed.
Finally our study supports the possibility of observing
precursor signals of the phase transition to a mixed hadron-quark matter
at high baryon density in the collision, central or semi-central, of
neutron-rich heavy ions in the energy range of a few $AGeV$.
As signatures we
suggest to look at the isospin structure of hadrons produced at high 
transverse momentum, as a good indicator of the neutron trapping effect. 
In conclusion the results presented here appear very promising for 
the possibility of exciting new results from dissipative collisions
with radioactive beams.

{\it Acknowledgements}

\noindent 
We warmly thanks A.Drago and A.Lavagno for the intense 
collaboration on the
mixed hadron-quark phase transition at high baryon and isospin density.



\end{document}